\documentclass[aps,pra,twocolumn,superscriptaddress]{revtex4}

\pdfoutput=1

\usepackage{amsmath,graphicx,amssymb,braket,xcolor,subfigure,upgreek,bbold,bm}
\usepackage[colorlinks, linkcolor=blue, citecolor=blue, urlcolor=blue, breaklinks=true]{hyperref}
\usepackage{verbatim}
\usepackage{float}
\usepackage{esint}
\usepackage{bm}
\usepackage{bbold}
\newcommand{\mi}{{\rm i}}
\newcommand{\me}{{\rm e}}
\newcommand{\id}{\mathbb{1}}

\begin{document}

\title{Dynamics of rotated spin states and magnetic ordering with two-component bosonic atoms in optical lattices}

\author{Araceli Venegas-Gomez}
\affiliation{Department of Physics and SUPA, University of Strathclyde, Glasgow G4 0NG, UK}
\author{Anton S. Buyskikh}
\affiliation{Department of Physics and SUPA, University of Strathclyde, Glasgow G4 0NG, UK}
\affiliation{Riverlane, Cambridge CB2 3BZ, UK}
\author{Johannes Schachenmayer}
\affiliation{CNRS, IPCMS (UMR 7504), ISIS (UMR 7006), and Universit\'{e} de Strasbourg, 67000 Strasbourg, France}
\author{Wolfgang Ketterle}
\affiliation{Harvard-MIT Center for Ultracold Atoms, Cambridge, Massachusetts 02138, USA}
\affiliation{Department of Physics, Massachusetts Institute of Technology, Cambridge, Massachusetts 02139, USA}
\author{Andrew J. Daley}
\affiliation{Department of Physics and SUPA, University of Strathclyde, Glasgow G4 0NG, UK}

\date{\today}

\begin{abstract}
The microscopic control available over cold atoms in optical lattices has opened new opportunities to study the properties of quantum spin models. While a lot of attention is focussed on experimentally realizing ground or thermal states via adiabatic loading, it would often be more straightforward to prepare specific simple product states and to probe the properties of interacting spins by observing their dynamics. We explore this possibility for spin-1/2 and spin-1 models that can be realized with bosons in optical lattices, and which exhibit \textit{XY}-ferromagnetic (or counterflow spin superfluid) phases. We consider the dynamics of initial spin-rotated states corresponding to a mean-field version of the phases of interest. Using matrix product state methods in one dimension, we compute both non-equilibrium dynamics and ground/thermal states for these systems. We compare and contrast their behaviour in terms of correlation functions and induced spin currents, which should be directly observable with current experimental techniques. We find that although spin correlations decay substantially at large distances and on long timescales, for induction of spin currents, the rotated states behave similarly to the ground states on experimentally observable timescales. 
\end{abstract}

\maketitle

\section{Introduction}

Control over cold atoms in optical lattices has led to opportunities to realize a range of spin-model Hamiltonians, arising from the superexchange of spinful fermions and bosons~\cite{RevModPhys.80.885,Bloch2012}. Important recent progress has been made in the realization of magnetically ordered states in such systems, with the observation of antiferromagnetic ordering of fermions, corresponding to a Heisenberg spin model~\cite{Parsons1253,Boll1257,Cheuk1260,Mazurenko2017,Mitra2017}. These states are generally produced by adiabatic loading of atoms into the lattice potential. In this context there have been a significant number of proposals for adiabatic manipulation of spin Hamiltonians in optical lattices, in order to obtain low-entropy states, even when the energy gap in the ground state is small~\cite{PhysRevLett.91.110403,PhysRevLett.96.250402,PhysRevLett.99.120404,PhysRevA.81.061603,PhysRevLett.107.165301,Schachenmayer2015}. This usually involves loading the lattice in a regime where the energy gap is large, and then manipulating the Hamiltonian parameters time-dependently. 

At the same time, recent experiments can prepare well-defined initial product states, which are not eigenstates of the system, and then probe their subsequent non-equilibrium dynamics~\cite{Greiner2002,J_Daley_2014}. Locally, these product states can appear as the mean-field state corresponding to the quantum phase associated with the ground state. For a particular phase, it is possible to directly probe in experiments to what extent the initial mean-field magnetic states, and the states they evolve into, are different from the true ground state for the same Hamiltonian parameters. For example, in the case of spin-1/2 models Barmettler et al.~\cite{Barmettler2009} considered the evolution of a perfect N{\'e}el state in one dimension (1D) under an antiferromagnetic Heisenberg Hamiltonian. As a result of the dynamics, the magnetic ordering is found to decay exponentially in time, thus demonstrating important differences between the mean-field and the true ground state in 1D.

In this article, we address such questions in the different context of spin-superfluid phases, which can be realized with multicomponent bosons in optical lattices ~\cite{Altman2003}. Such spin-superfluids can also be identified with an \textit{XY}-ferromagnet, and proposals for their adiabatic state preparation have been discussed, especially for the spin-1 case that occurs with two particles per site ~\cite{Schachenmayer2015}. On the other hand, an ideal \textit{XY}-ferromagnetic state can also be well approximated by a mean-field description, where all of the spins point in the \textit{XY} plane. For large spins, this corresponds to approximating the spin superfluid by a product of spin coherent states, analogously to a superfluid state of bosons on a lattice~\cite{JakschZoller2005,PhysRevA.83.043614}. Moreover, such states can be prepared in a relatively straight-forward experimental sequence. Beginning in a Mott Insulator (MI) state in which all spins are initially prepared aligned along the $z$-axis, we can apply an rf transition to rotate the state into the \textit{XY} plane, and we call this state the rotated state. 

We compare and contrast exact quantum ground states of spin-1/2 and spin-1 models to their rotated-spin (mean-field) counterparts. Focusing on the 1D case, we compute ground and thermal states as well as the many-body dynamics of the system using tensor network methods based on Matrix Product State (MPS) and Matrix Product Operator (MPO) techniques~\cite{White1992, White1993,Daley2004,Verstraete2004,SCHOLLWOCK201196}. We first quantify how far the initial spin coherent mean-field state is from the true \textit{XY}-ferromagnetic ground state. We then study the evolution of this state, which is a consequence of inter-species interactions (anisotropies in the effective spin-models) leading to the initial state not being an exact eigenstate of the Hamiltonian. We find that for short times ``ideal'' \textit{XY}-correlations remain relatively robust. For longer times and small anisotropies the dynamics produces states with exponentially decaying spin correlations, resembling thermal states. We analyze the dependence of correlation lengths on anisotropies. For large anisotropies, the thermalization picture breaks down and a non-equilibrium state very different from the ferromagnet builds up quickly.
Lastly, we show how the effective magnetic ordering, i.e.~spin superfluidity can be probed by inducing spin currents. We propose a way to probe the magnetic ordering by measuring spin-currents generated by an effective magnetic field gradient. We compare spin-currents following from true ground-states of the system and for initial spin coherent states.

The remainder of this article is organized as follows:
In Sec.~\ref{sec:models}, we review the two effective spin models we consider in this work (spin-1 and spin-1/2), and how they arise from a two-species Bose-Hubbard model. In Sec.~\ref{sec:out-of-eq_synamics}, we explore the differences between ground, rotated, and thermal states in out-of-equilibrium dynamics. In Sec.~\ref{sec:probing_state} we discuss methods to probe these states by observing spin currents of bosons in an optical lattice. Lastly, we provide a summary and an outlook in Sec.~\ref{sec:summary}.

\section{Spin Models from two-component bosons}
\label{sec:models}

In this section, we introduce the two effective spin models that we will analyze in this work. Both can appear as effective models in Mott Insulating states of two-components (e.g.~two internal spin states) of bosons trapped in the lowest band of an optical lattice~\cite{Duan2003,Kuklov2003}, and such systems exhibit a rich ground-state phase-diagram~\cite{Altman2003}. The system is described by a two-species Bose-Hubbard Hamiltonian
\begin{equation}
\label{eq:2species_BH}
\begin{split}
\hat{\mathcal{H}} = & -\zeta \sum_{\langle i,j \rangle} (\hat{a}_i^\dagger \hat{a}_j + \hat{b}_i^\dagger \hat{b}_j) + U_{AB} \sum_j \hat{a}_j^\dagger \hat{a}_j \hat{b}_j^\dagger \hat{b}_j \\
& + \frac{U_A}{2} \sum_j \hat{a}_j^\dagger \hat{a}_j^\dagger \hat{a}_j \hat{a}_j + \frac{U_B}{2} \sum_j \hat{b}_j^\dagger \hat{b}_j^\dagger \hat{b}_j \hat{b}_j.
\end{split}
\end{equation}
Here, $\hat{a}_i$ and $\hat{b}_i$ are the bosonic annihilation operators for the two species denoted as $A$ and $B$, respectively. The notation $\langle i,j \rangle$ denotes a sum over all nearest-neighbour sites, $\zeta$ is the tunneling rate, and $U_A,U_B$ the intra-species and $U_{AB}$ the inter-species on-site interaction energy strength. We denote the average occupation of particles per site as $n$. We will consider equal intra-species interactions $U \equiv U_A = U_B$, which describes very well the situation for $^{87}$Rb atoms.

In the case of integer $n$, when the intra-species interactions are large compared with the tunnelling, $U_A, U_B \gg \zeta$, the ground-state of the model is a MI state with particles exponentially localized at each lattice site and with small local number fluctuations. In second order perturbation theory, analogous to a Schrieffer-Wolf transformation producing the Heisenberg model from the Hubbard model~\cite{Altman2003}, we obtain an effective spin-Hamiltonian acting in the low energy subspace.

\begin{figure}[tb]
\centering
\includegraphics[width=\columnwidth]{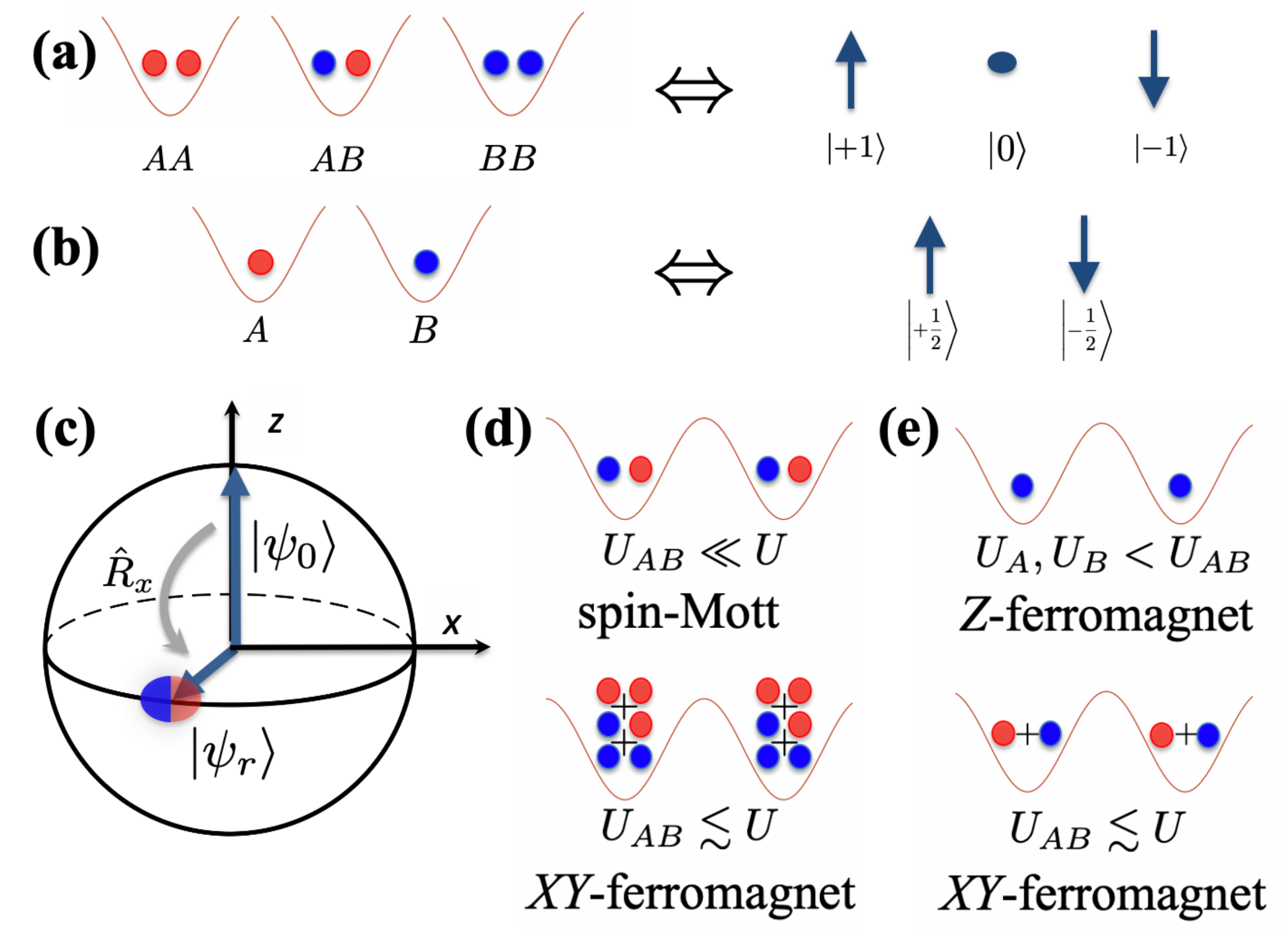}
\caption{Effective spin models for two-component bosons. (a/b) Correspondence of local particle states to spin-states for a spin-1 model and a spin-1/2 model, respectively. (c) Rotated state: A $\pi/2$ spin rotation around the $x$-axis is applied to a state with all spins initially aligned along the $z$-axis. The rotated states $\ket {\psi_r} $ are superpositions of the different local particle states. (d/e) Sketch of local superpositions of particle states, corresponding to the mean-field picture of the ground-state phases in different regimes for the spin-1 and spin-1/2 model, respectively.}
\label{fig:Overview_Figure}
\end{figure}

\subsection{Spin-1 Model}
For $n=2$ the low-energy sub-space on a site $l$ can be represented by three different states $\ket{+1}_l$, $\ket{0}_l$, $\ket{-1}_l$, as depicted in Fig.~\ref{fig:Overview_Figure}(a), comprising effective eigenstates of a diagonal spin-1 operator, $\hat{S}^z_l$, with eigenvalues $S^z_l=+1,0,-1$. The effective spin states correspond to the respective particle states
$\hat{a}_l^\dagger \hat{a}_l^\dagger \ket{0}$,
$\hat{a}_l^\dagger \hat{b}_l^\dagger \ket{0}$, and 
$\hat{b}_l^\dagger \hat{b}_l^\dagger \ket{0}$, where $\ket{0}$ denotes the empty lattice state.

Considering the case of an equal number of $A$ and $B$ bosons ($n_A=n_B$), the effective Hamiltonian is an anisotropic spin-1 Heisenberg model \cite{Altman2003},
\begin{equation}
\label{eq:Spin1-H}
\hat{\mathcal{H}}_\text{SP1} = - J \sum_{\langle i,j \rangle} \bold{\hat{S}}_i \cdot \bold{\hat{S}}_j + u \sum_j (\hat{S}^z_j)^2.
\end{equation}
Here,
$u = U - U_{AB}$, $J = 4 \zeta^2 / U_{AB}$, and $\hat{\bold{S}}_i = (\hat{S}^x_i, \hat{S}^y_i, \hat{S}^z_i)$ is a vector of the three spin-1 operators.

 The ground-state phase diagram of Eq.\eqref{eq:Spin1-H} has been studied in \cite{Chen2003}. The magnetic ordering in the ground state depends on the interactions. When $U \gg U_{AB}$, the ground state will exhibit a spin insulator or spin-Mott state configuration, with $S_i^z\rightarrow 0$ for all sites $i$.
Interactions of similar size, $U_{AB} \lesssim U$, lead to a \textit{XY}-ferromagnetic ground state, induced by the superexchange term. The rotated product-state, which would represent a mean-field \textit{XY}-ferromagnetic state is a superposition of all three spin states on each site (as sketched in Fig.~\ref{fig:Overview_Figure}(d)). 
Note that, while in the Mott phase of two species of atoms the net overall transport of atoms is suppressed, the \textit{XY} phase corresponds to a state with a counterflow, (i.e.~the currents of the two species are equal in absolute values but opposite directions), and can be nondissipative (supercounterflow) \cite{Kuklov2003,RevModPhys.80.885}. Finally, for $U_{AB}>U$, the ground state is a $z$-ferromagnet.

\subsection{Spin-1/2 Model}

In the case of $n=1$, the resulting effective Hamiltonian is a spin-1/2 XXZ Heisenberg model, where a single $A$ boson is mapped to spin-up $\ket{\uparrow}_j$ and $B$ boson to spin-down $\ket{\downarrow}_j$ on site $j$ (cf.~Fig.~\ref{fig:Overview_Figure}(b)) ~\cite{Altman2003}
\begin{equation}
\label{eq:XYFM_Hamiltonian}
\hat{\mathcal{H}}_{\text{SP1/2}}= - 
J \sum_{\langle i,j \rangle} \hat{\bm{\sigma}}_i \cdot \hat{\bm{\sigma}}_j + \Delta \sum_{\langle i,j \rangle} \hat{\sigma}^z_i  \hat{\sigma}^z_j,
\end{equation}
where $J = 4 \zeta^2 / U_{AB}$ and $\Delta = 8 \zeta^2 / U_{AB} - 8 \zeta^2 / U$ is the anisotropy. $\hat{\bm{\sigma}}_i = (\hat{\sigma}^x_i, \hat{\sigma}^y_i, \hat{\sigma}^z_i)$ is a vector of the three Pauli matrices. 

We note that in an experiment, variations in $\Delta/J$ correspond to variations of $u=U-U_{AB}$, as defined in the spin-1 case, which we can rewrite as $\Delta/J = 2u/U$. The realisable range of $\Delta/J$ values is thus dependent on our ability to tune $U_{AB}$ in an experiment.


In the spin-1/2 case, a phase transition occurs at $U_{AB}=U$, i.e. for $\Delta = 0$. When $U_{AB}\lesssim U$ ($\Delta \gtrsim 0$), the ground state of the system is \textit{XY}-ferromagnetic (or spin superfluid), in contrast to the \textit{Z}-ferromagnet for $U_{AB}>U$ ($\Delta <0$) [see Fig.~\ref{fig:Overview_Figure}(e)]. Note that for $U_{AB} = U/2$ ($\Delta = J$) the Ising coupling vanishes, i.e.~the model becomes equivalent to that of non-interacting hard-core bosons (or non-interacting fermions), and for $U_{AB} < U/2$ ($\Delta > J$), the sign in front of the Ising coupling term becomes negative (anti-ferromagnetic). In this work we focus on the regime of $0 \leq \Delta \leq 2J$, for which the true ground-state exhibits quasi-long-range order.

Note that in 1D, the spin-1/2 XXZ Heisenberg model have been extensively studied (see e.g.~\cite{Barmettler_Quant_2010}). Note that it is generally integrable, i.e.~it can be diagonalized by a Bethe ansatz solution, which can lead to certain exact solutions for simple observables in equilibrium. For the situation considered here, the XXZ model is gapless in the thermodynamic limit and can be described by a Luttinger model. More generally, universal valid predictions on correlation dynamics after quenches in the gapless phase of the spin-1/2 XXZ Heisenberg model (as studied below) have been made from conformal field theories (CFT)~\cite{Calabrese_Time_2006,Calabrese_Quant_2016}, even though such theories are technically valid only for low-energy quenches and in the thermodynamic limit.

\section{Rotated states and out-of-equilibrium dynamics}
\label{sec:out-of-eq_synamics}

In this section, we will first discuss the preparation of the spin-rotated states, and then study their differences to true \textit{XY}-ferromagnetic ground states. Second, we will look at the dynamics of the system and analyze the dynamically prepared states, e.g.~as a function of the anisotropies.

\subsection{Preparation of spin rotated states}

The ideal mean-field \textit{XY}-ferromagnetic state can be prepared by beginning with all spins aligned along the $z$-axis ($ \ket{\psi_0} = \ket{\uparrow\uparrow\uparrow\uparrow\uparrow\uparrow\uparrow...}$), and then rotating that state locally into the \textit{XY} plane, of every atom simultaneously (see Fig.~\ref{fig:Overview_Figure}(c)). In an experiment, this could be achieved by beginning in a single-component MI with the correct filling factor, and then applying an rf/microwave drive that corresponds to a $\pi/2$ rotation around the $x$-axis (see e.g.~\cite{Hild_Far-f_2014}), as generated by the operator
\begin{equation}
\hat{R}_x = \prod_j \me^{-\mi \frac{\pi}{2} \hat{S}^x_j}.
\end{equation}
The rotated initial state is then
\begin{equation}
\ket{\psi_r} = \hat{R}_x \ket{\psi_0}.
\end{equation}
The operation is calculated analogously for the spin-1/2 with operators $\hat{\sigma}^x$.
We will now analyze how close the state $\ket{\psi_r}$ is to the true \textit{XY}-ferromagnetic ground-state and how the dynamics will modify it.

\subsection{Comparison of the rotated state with the ground state}

\begin{figure}[tb] 
\includegraphics[width=\columnwidth]{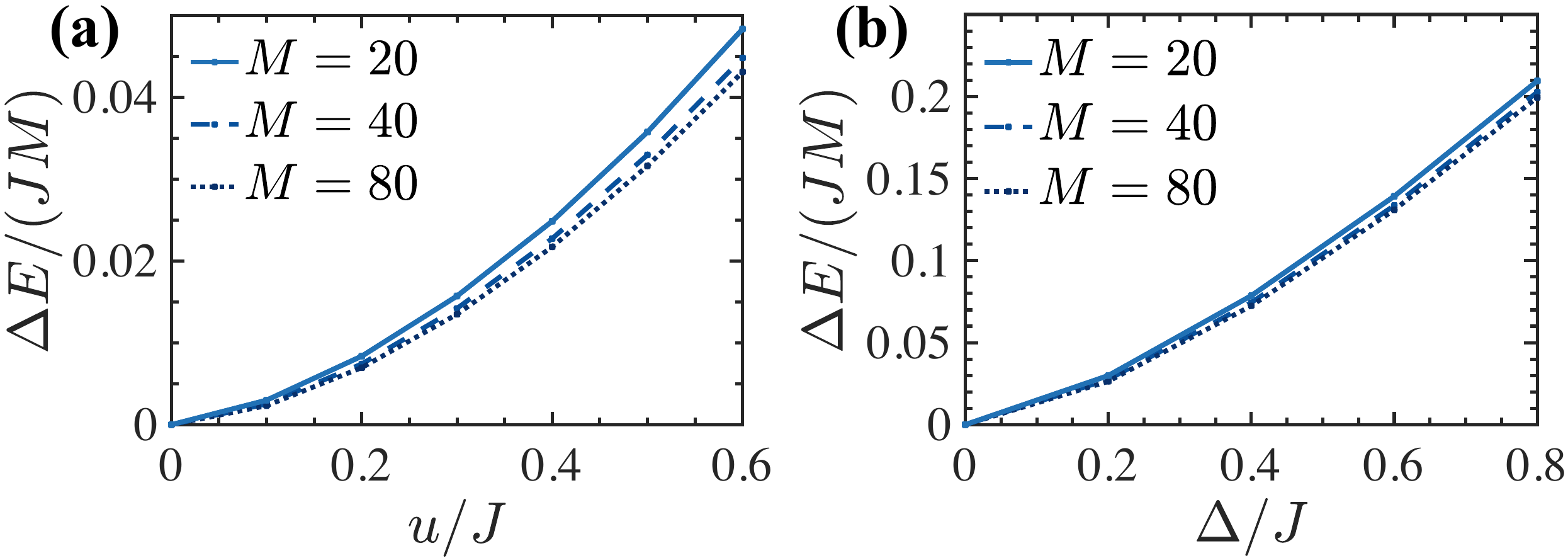}
\caption{
Energy difference per particle between the reference state (rotated state) and the ground state of the Hamiltonian, for different systems sizes $M$ and different anisotropy $u/J,\Delta/J$, for the effective (a) spin-1 and (b) spin-1/2 models of two-component bosons on an optical lattice.
The bond dimension used for the MPS calculations was $D=64$, with open boundary conditions.}
\label{fig:Spin_Ediff}
\end{figure}

To obtain a first idea about the similarities of $\ket{\psi_r}$ and the ground-state of the system, we compute the energy difference $\Delta E = E_r - E_{GS}$ per spin of the two states. Fig.~\ref{fig:Spin_Ediff} shows $\Delta E$ for various values of the anisotropy ($u/J$ and $\Delta/J$) for spin-1 and spin-1/2, respectively. The results are for a 1D chain with a varying number of sites $M$, and computed using MPS techniques. There is no energy difference without anisotropy, and the result is very close to the ground state for small values of $u/J,\Delta/J$ in each case. The energy difference increases with the anisotropy, and for large system sizes the value of energy difference per spin is independent of the system size.

It is difficult to compare these differences directly between the two models, because the Hamiltonians and their corresponding energy scales are significantly different. We also note that the variation in $U-U_{AB}$ that is required for a change in $\Delta/J$ for the spin-1/2 model is much larger than the variation for a given value of $u/J$ in the spin-1 model, as for the Mott Insulator regime in which we are working, $U/J$ is substantial.

Naturally, the energy difference only gives us a first indication of similarities or differences between the rotated state and the ground states. In the next sections we will look at the time evolution of correlation functions, and then the behaviour of spin currents induced in the system. 

\subsection{Dynamics of correlation functions}

We now look at the out-of-equilibrium dynamics after a preparation of $\ket{\psi_r}$, in particular we will focus on the dynamics of spin-spin correlations in the system.
To compute the time evolution under each Hamiltonian, we use the Time Evolving Block Decimation (TEBD) algorithm \cite{Vidal2003, Vidal2004, Verstraete2004,SCHOLLWOCK201196} for MPS. The corresponding bond dimensions required for convergence are indicated in the figure captions.

The correlation functions are calculated as
\begin{equation}
\Theta_j = | \overline{\braket{S_i^+ S_{i+j}^-}} | = \frac{1}{M-2b-j} \sum^{M-b-j}_{i=1+b} | \braket{\hat{S}^+_i \hat{S}^-_{i+j}} |,
\label{eq:Correlations_Expression}
\end{equation}
where $i$ denotes the index of the site, $j$ is the distance or number of sites, and $b=M/5$ is a number of sites at the boundary that we omit to reduce the open boundary effects. The correlations are calculated analogously for the spin-1/2 with operators $\hat{\sigma}^+_i,\hat{\sigma}^-_{i+j}$.

\begin{figure}[tb] 
\includegraphics[width=\columnwidth]{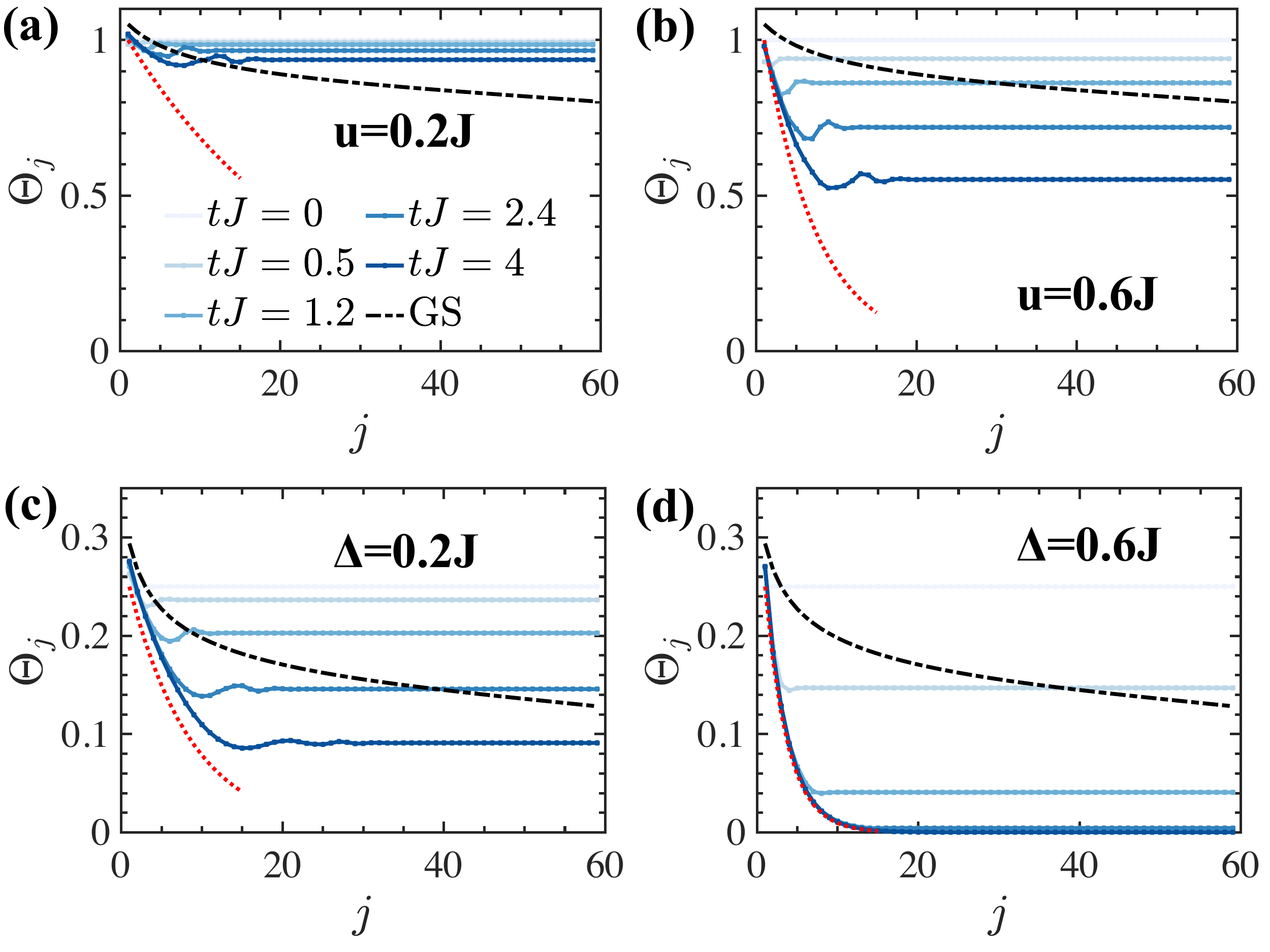}
\caption{
Comparison of the decay of the correlations with distance at different snapshots in the time evolution ($tJ=0$, $tJ=0.5$, $tJ=1.2$, $tJ=2.4$, $tJ=4$).
The black dash-dotted line indicates the value of the correlations for the ground state (GS) of the corresponding Hamiltonian. Results are for the spin-1 (a,b) and spin-1/2 (c,d) model, respectively.
The different panels contrast the evolution for different anisotropies.
The red dotted lines indicate the exponential decay of the corresponding thermal states with the correlation lengths calculated in Fig.~\ref{fig:Thermal_Analysis}.
[The calculations were performed for a system size $M=100$, bond dimension for the MPS calculations $D=128$ for spin-1 and $D=256$ for spin-1/2, and open boundary conditions.]}
\label{fig:CorrelationEvol_Distance}
\end{figure}

In Fig.~\ref{fig:CorrelationEvol_Distance} we show the correlations at different times, and compare them with correlations of the corresponding ground-state. In both models, the correlations for the rotated state begin at a larger value at long distances because of the choice of initial state, but then decay rapidly in time, especially at long distances. This decay is clearly faster for increased anisotropy in both models, and is especially rapid in the spin-1/2 model for $\Delta=0.6J$. In the spin-1 case, for $u=0.2J$ the decay of the correlations at $tJ=4$ is minimal, indicating that for a small value of the anisotropy $u/J$ the magnetic order remains relatively robust under time evolution. For $u=0.6J$, in contrast, the correlations decay faster with time, but still conserve the magnetic ordering at long distances. In the spin-1/2 case, even for a small value of the anisotropy $\Delta=0.2J$ we can see how the correlations decrease quickly on the time-scale of a few tunneling times. For $\Delta=0.6J$, the spin-ordering vanishes rapidly to zero. 

\begin{figure}[tb] 
\includegraphics[width=\columnwidth]{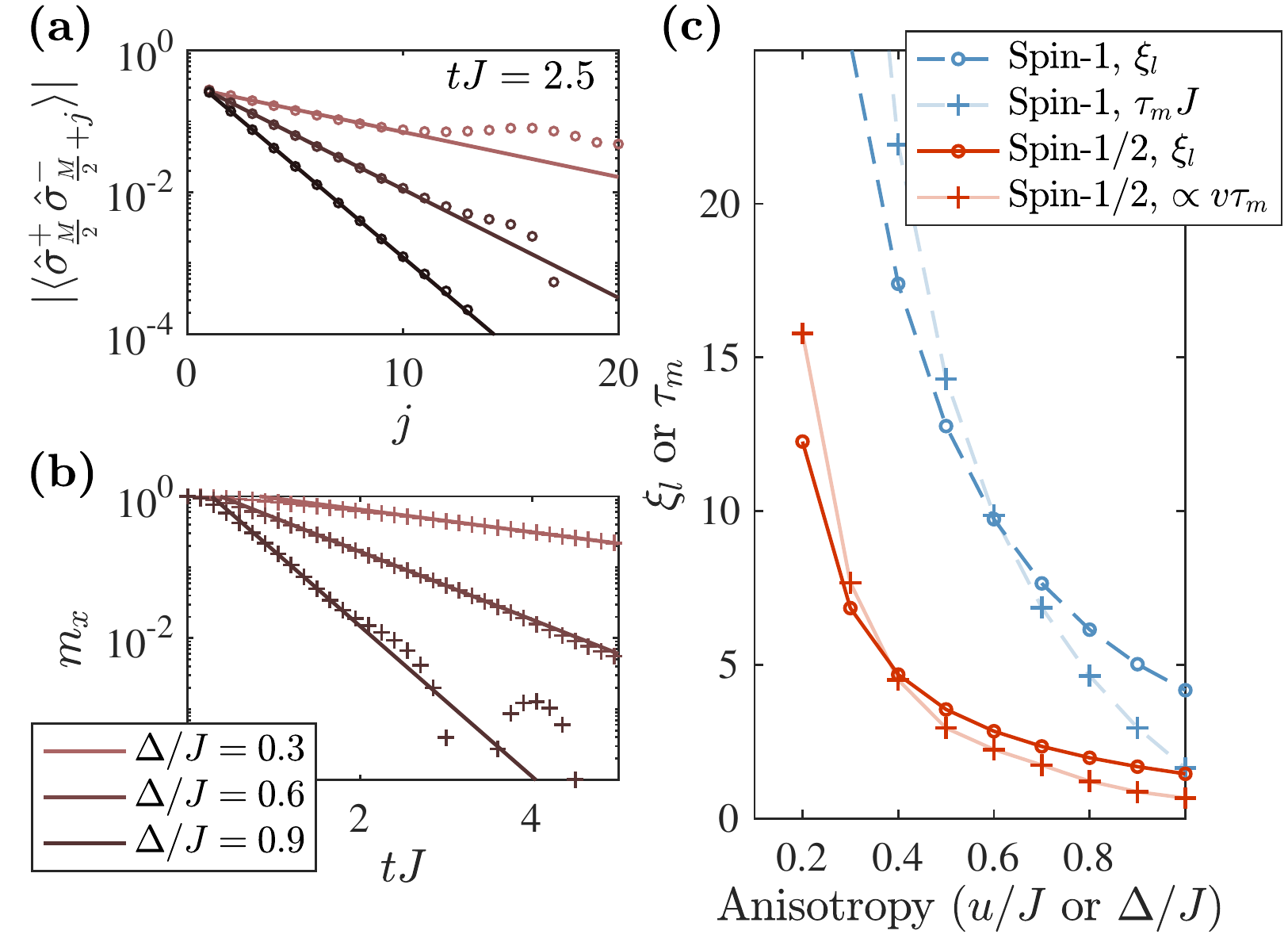}
\caption{Example fits to the exponential decay of (a) correlations at short distances within the light cone (at $tJ = 2.5$), and (b) magnetization as function of time (at late times, but before boundary effects) in the spin-1/2 model. (c) Scaling of the fitted correlation lengths ($\xi_l$) and the  magnetization decay time ($\tau_m$, scaled) from (a) and (b) as function of the anisotropy $\Delta$. Results are also shown for the spin-1 model as function of $u$. [$M=40$, MPS bond dimension $D=256$, open boundaries.]}
\label{fig:CFT}
\end{figure}

The observed dynamics in the decay of correlations in Fig.~\ref{fig:CorrelationEvol_Distance} is consistent with a usual light-cone spreading of entangled quasi-particle excitations, as is predicted for the spin-1/2 model, e.g., by CFT~\cite{Calabrese_Quant_2016}. In particular, for short distances, within the light-cone (i.e.~for distances $|i-j| < v_m t$, with $v_m$ the maximum velocity of entangled excitations spreading through the chain), one expects $|\langle \hat \sigma_i^+ \hat \sigma_j^-  \rangle| \propto {\exp}(-(|i-j|/\xi_l)$ with some characteristic correlation length $\xi_l$. This spatial correlation decay is expected to be related to the time-dependent decay of local observables, such as the magnetization, which should decay as $m_x(t) = (1/M) \sum_i \langle \hat \sigma_i^x    \rangle  \propto \exp(-t/\tau_m)$ on some characteristic time-scale $\tau_m \propto \xi_l$. We test those predictions in Fig.~\ref{fig:CFT} for our simulations in a $M=40$ site system and an initial state polarized along the $x$ direction. Note that our scenario is different from the case of an initial N\'eel state considered e.g.~in \cite{Barmettler_Quant_2010}. While for the N\'eel state the staggered $z$ magnetization decays slower when increasing $\Delta$ (i.e.~when moving away from the isotropic point), here the $x$ magnetization is conserved at the isotropic point and decays faster with increasing $\Delta$. 

In agreement with the CFT prediction, in Fig.~\ref{fig:CFT} (a) and (b), we observe exponential decay of both $|\langle \hat \sigma_i^+ \hat \sigma_j^-  \rangle|$ in space at short distances, and for $m_x(t)$ in time (for sufficiently long times, before boundary effects become important, typically at $tJ\sim 2.5$ for $\Delta \sim J$). In Fig.~\ref{fig:CFT}(c) we compare the scaling of $\xi_l$ and $\tau_m$ as function of $\Delta$. We expect $\xi_l \propto v t$ with a velocity proportional to $v = \sqrt{2 \Delta - \Delta^2}/\arccos(1- \Delta)$ \cite{Giamarchi_Quant_2003}. Indeed we observe a very similar scaling of  both quantities $\tau_m$ and $\xi_l$. The slightly stronger dependence on $\Delta$ of $\tau_m$ than $\xi_l$  may be attributed either to the difficult fitting procedure in the relatively small system considered here (especially in the limits of large and small $\Delta$), or to the fact that our setup is outside of the expected validity of the CFT approach, as our quench produces a high energy state in the middle of the many-body energy spectrum~\cite{Calabrese_Quant_2016}.  

We also repeated the same analysis for the non-integrable spin-1 model and results for the scaling of $\tau_m$ and $\xi_l$ are also shown in Fig.~\ref{fig:CFT}(c). Interestingly, we find a very similar behavior as in the spin-1/2 case, with a difference between the scaling of $\tau_m$ and $\xi_l$. Note that the observed exponential decay of correlations within the light-cone is also consistent with thermal states [see red dashed lines Fig.~\ref{fig:CorrelationEvol_Distance}]. It is quite surprising that this thermal behavior at short distances is more pronounced in the integrable spin-1/2 case compared to our simulations in the non-integrable spin-1 model. 
This makes such setups interesting for possible larger-scale experimental tests. We will analyze the thermalization behavior in the following sections in more detail.

\subsection{Thermal states} 

In this section, we now compare the dynamically obtained states to a thermal state $\hat \rho_{\rm th} \propto {\exp(-\beta \hat H})$ with an inverse temperature $\beta=1/T$ ($k_B \equiv 1$) such that the energy of the thermal state matches the energy of the rotated state,
\begin{equation}\label{eq:S_Tinf}
\braket{E}_\beta = {\rm tr}(\hat \rho_{\rm th} \hat H) = E_r.
\end{equation}
The state $\hat \rho_{\rm th}$ describes the system in the long-time limit effectively for simple observables if it thermalizes, e.g.~for non-integrable models in the absence of localization \cite{Deutsch1991,Srednicki_Chaosa_1994,Rigol2008, Deutsch_Eigenst_2018, DAlessio_Fromqu_2016,Eisert2015a, Nandkishore_Many-Bo_2015}.
In order to compute properties of the system (for large system sizes) at finite temperatures, we use an imaginary-time evolution of the density matrix in MPO form \cite{Verstraete2004}. At the initial point of the evolution the system is considered at the infinite temperature, i.e. its density matrix is proportional to the identity, $\rho_0\propto \id$, where all states have equal probability of occupation. Then, the next step is to evolve the density matrix to finite temperatures $\hat \rho(\beta) \propto \me^{-\beta\hat{H}}$. We use a purification technique~\cite{Verstraete2004, Cuevas2013} to preserve positive semi-definiteness of the density matrix, and hence rewrite this expression as $\hat \rho(\beta) \propto \me^{-\beta \hat H/2} \rho_0 \me^{-\beta \hat H/2}$. Since $\rho_0 = \rho_0^2 = \rho_0 \rho_0^\dagger$ only one side of the above expression needs to be evolved, $\bar{\rho}(\beta) \equiv \me^{-\beta \hat H/2} \rho_0$, and the thermal expectation value of an arbitrary operator $\hat{O}$ can be obtained as
\begin{equation}
\braket{\hat{O}}_\beta = \frac{\mathrm{tr}[\hat{O}\bar{\rho}(\beta)\bar{\rho}^\dagger(\beta)]}{\mathrm{tr}[\bar{\rho}(\beta)\bar{\rho}^\dagger(\beta)]}.
\end{equation}

We find that the Time-Dependent Variational Principle (TDVP) algorithm \cite{Haegeman2011, Koffel2012, Haegeman2013, Haegeman2016} is a very efficient integrator for time-propagation at finite temperatures in terms of the balance between the speed and accuracy, however other methods maybe useful in terms of accuracy, for instance Runge-Kutta \cite{Ascher1998}.

The accuracy of the method is verified by comparing the numerical calculations with the exact solution (using exact diagonalization) in the case of smaller systems.
For bigger system sizes the convergence of numerical results to the exact solution is checked by increasing the MPO/MPS bond dimension, $D$. We verified the validity of all our results by comparing the convergence of this method with respect to the observables we are interested in, and we confirm the convergence in the bond dimension by running multiple calculations with increasingly large $D$.

\begin{figure}[tb] 
\includegraphics[width=\columnwidth]{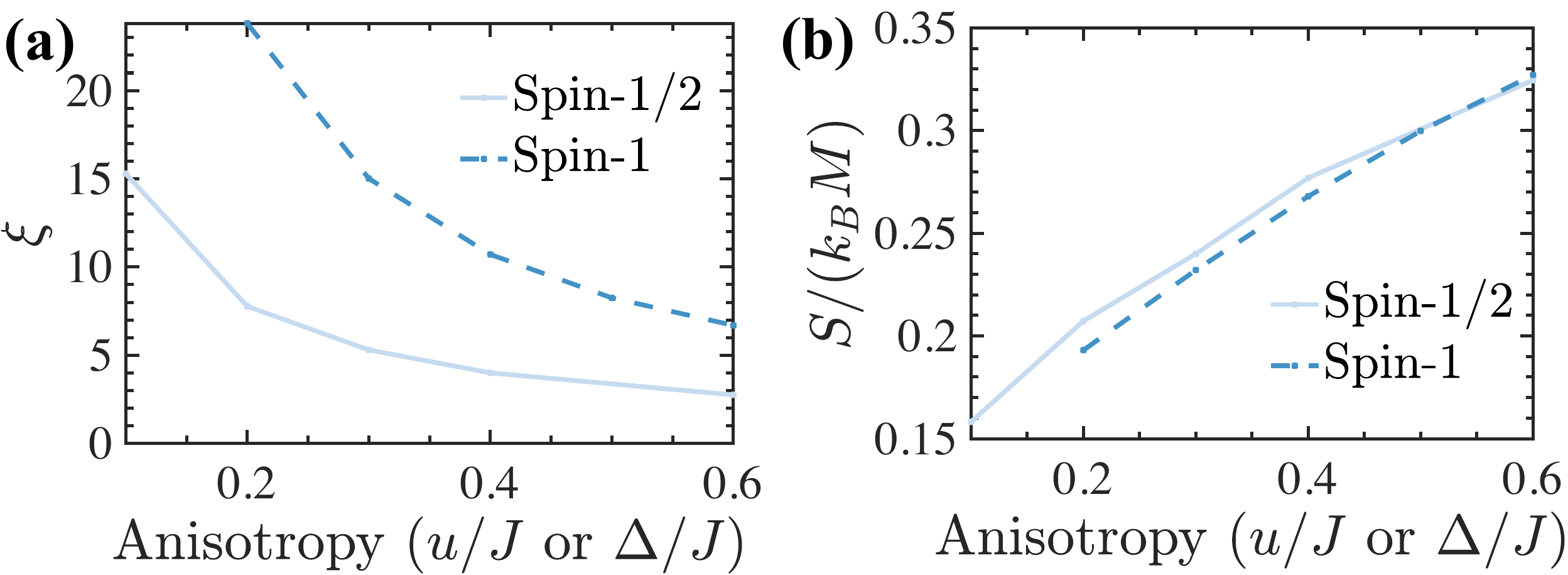}
\caption{
Properties of thermal states with identical energy as the rotated state for both the spin-1 and spin-1/2 model.
(a) Correlation length $\xi$ obtained from an exponential fit to the decay of $\braket{\hat S^+_i \hat S^-_{i+r}}$ with $r$ \eqref{eq:CorrelationEq}.
Shown is $\xi$ as a function of the anisotropies $u/J$ and $\Delta/J$. Here, $M = 40$, correlations calculated as in \eqref{eq:Correlations_Expression}.
(b) Entropy per particle, $S/(k_B M)$, as function of the anisotropies, $M=40$.
The bond dimension used for these MPS calculations was $D=64$, with open boundaries.}
\label{fig:Thermal_Analysis}
\end{figure}

In the ground state of our spin models, we see that the correlations decay algebraically (as shown by the black lines in Fig.~\ref{fig:CorrelationEvol_Distance}). However, this effect will be destroyed by thermal effects: at high temperatures, spin orientations become randomized, and the correlations will exponentially decrease with increasing distance $r$,
\begin{equation}\label{eq:CorrelationEq}
\braket{\hat S^+_i \hat S^-_{i+r}} \propto \me^{-\frac{r}{\xi(T)}},
\end{equation}
with $\xi$ being the correlation length (analogously for the spin-1/2 with operators $\hat{\sigma}^+_i, \hat{\sigma}^-_{i+r}$). The properties of the thermal states corresponding to the energies of the rotated states are summarized in Fig.~\ref{fig:Thermal_Analysis}. There we compare results for the correlation lengths (obtained from an exponential fit), and for the entropy per lattice site.

In both models the correlation length decreases with the anisotropy. For the spin-1 model, we find that a large correlation length is attained for smaller anisotropy, demonstrating that for a thermal state (in the long-time limit) a state with significant correlations may be stabilized for small $u/J$. In contrast, for the spin-1/2 case we find that except for very small $\Delta/J$ the correlation lengths obtained are shorter. 
 
Note that in performing an effective imaginary time evolution as described above, the exponential factor will always provide an instability towards the ground state, where numerical noise biases the final state towards the ground state, especially for large $\beta$. Thus, in our calculation for thermal states, the calculations become inaccurate in the low-temperature limit. For the spin-1 model, this limited our comparisons to the regime $u \gtrsim 0.2J$, where the correlation length for the spin-1 model becomes comparable to the system size. 


\subsection{Thermalization dynamics} 

The relaxation and comparison with thermal states in the previous section can be extended to consider to what extent local observables relax to values we might expect for corresponding thermal states. In general, we expect that closed quantum systems will thermalize in the long-time limit in the sense that local observables in a small subsystem appear to be described by a thermal density matrix $\hat \rho_\text{th} \propto \exp(-\beta \hat H)$, with the (inverse) temperature set by the energy matching condition with the initial state, $\bra{\psi_0}\hat H \ket{\psi_0} = \text{tr}(\hat \rho_\text{th} \hat H)$. This thermalization behavior is expected for Hamiltonians without simple/local conserved degrees of freedom (integrable models) and in situations without disorder. The mechanism behind such thermalization can be analysed, e.g., via the eigenstate thermalization hypothesis~\cite{Deutsch1991,Srednicki_Chaosa_1994,Rigol2008, Deutsch_Eigenst_2018, DAlessio_Fromqu_2016,Eisert2015a, Nandkishore_Many-Bo_2015}. 

Here we ask to what extent local observables relax towards thermal values on short timescales in different parameter regimes. Ideally, a comparison would be made at long times, but we are naturally limited by the ability to compute time-dependent dynamics of large systems (for methods based on tensor networks, this is related to the growth of entanglement entropy \cite{SCHOLLWOCK201196}). To avoid finite-size effects of the whole system, we consider the time dependent expectation value of the local subsystems of $M=40$ spins, specifically choosing an example observable $\braket{\hat{O}_j}=\braket{\hat{S}^+_i \hat{S}^-_{i+j}}$ (analogously for the spin-1/2 with operators $\hat{\sigma}^+_i, \hat{\sigma}^-_{i+j}$), and perform a system average as in Eq.~\eqref{eq:Correlations_Expression}. 

\begin{figure}[tb] 
\includegraphics[width=\columnwidth]{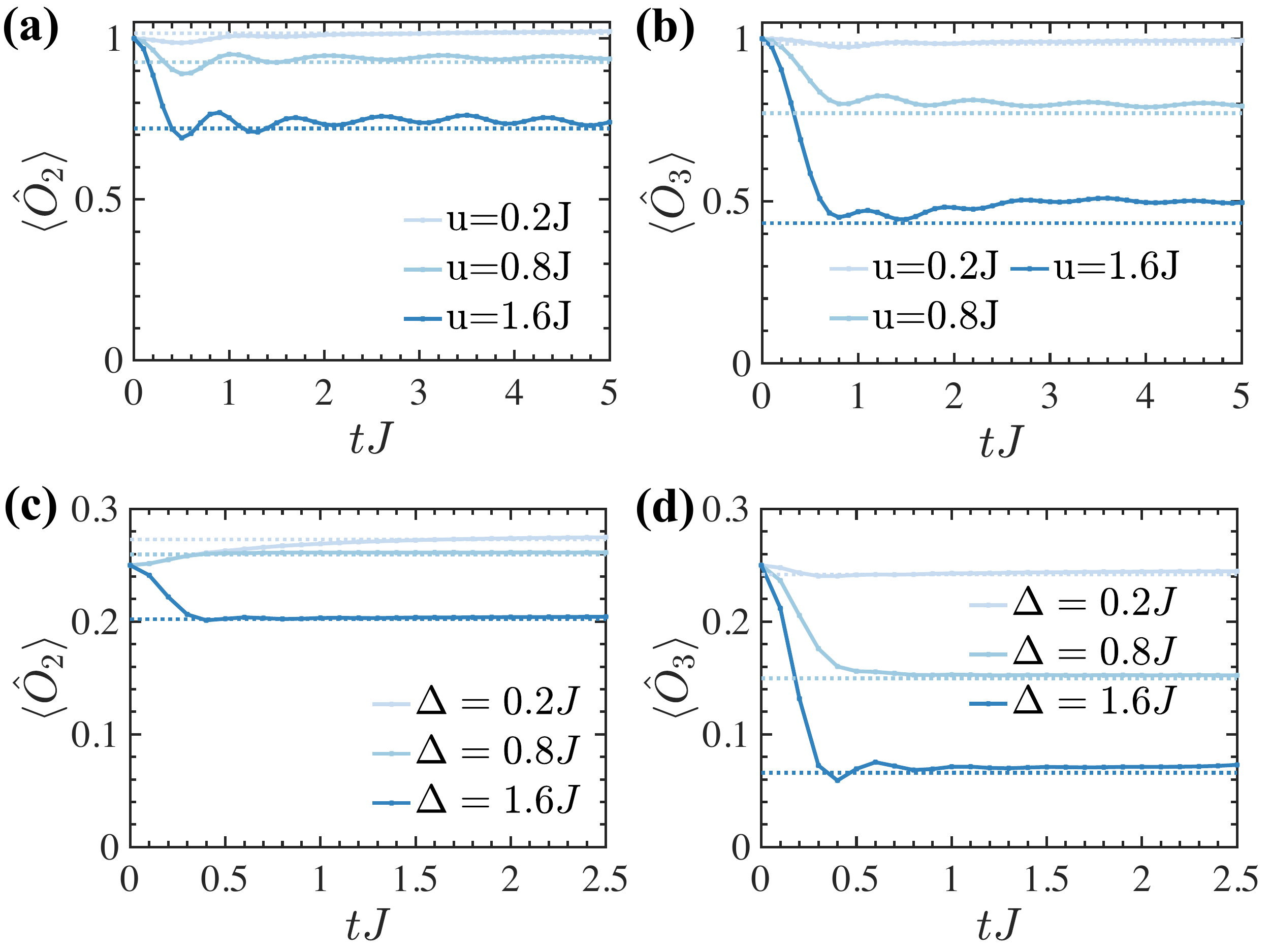}
\caption{Thermalization behavior of correlations, for both the spin-1 (a,b) and spin-1/2 (c,d) model. The figures show correlation functions in comparison to those of expected thermal states of energy $E_r$ (dotted lines), as a function of time. As observable we consider the correlations $\braket{\hat{O}_j}$ for different distances $j=2$ in panels (a,c) and $j=3$ in panels (b,d), always with $b=18$. [The calculations were performed for a system size $M=40$ and system averaged as in Eq.~\eqref{eq:Correlations_Expression}, bond dimension for the MPS time evolution calculations $D=128$ for spin-1 and $D=256$ for spin-1/2, and open boundary conditions.]}
\label{fig:SP_ETH_Time}
\end{figure}

In Fig.~\ref{fig:SP_ETH_Time} we show the relaxation behaviour of these correlations, both for the spin-1/2 and the spin-1 models, and for correlations at a separation of two sites and three sites. In each case, we show the equivalent values of the correlations from thermal states with the same mean energy. We note that in the spin-1/2 model, the spins relax on a short timescale to values close to the equivalent thermal value. For the spin-1 case, the values reached are also close, but quantitative agreement becomes worse for larger values of $u$ and larger separations.

\begin{figure}[tb] 
\includegraphics[width=7cm]{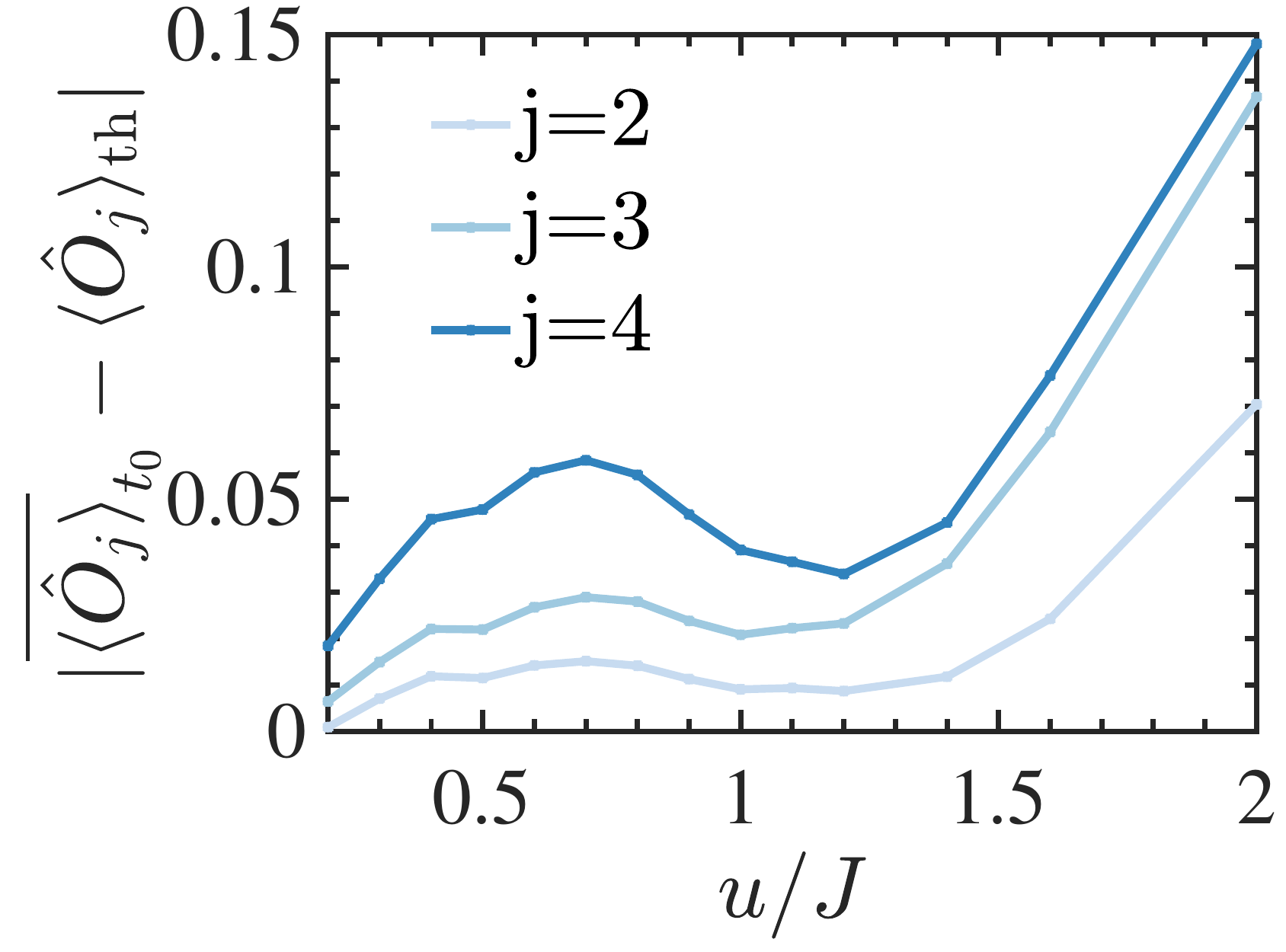}
\caption{Thermalization behavior of correlations for the spin-1 model. The figure show differences of time-averaged correlation functions ($\hat{O}_j$, for different distances $j$, and $b=18$) to those of expected thermal states of energy $E_r$, where a value of zero indicates dynamics towards a thermal state. This difference is shown for a long-time average $t_0 J\in[2,5]$, and as function of the anisotropies. [The calculations were performed for a system size $M=40$ and system averaged as in Eq.~\eqref{eq:Correlations_Expression}, bond dimension for the MPS time evolution calculations $D=128$, and open boundary conditions.]}
\label{fig:SP_ETH}
\end{figure}

To summarise the results for different parameter regimes, we can compute $\braket{\hat{O}_j}$ averaged over a time-scale $t_0 \in [t_{\rm{in}},t_{\rm{fin}}]$, $\overline{\braket{\hat{O}_{j}}}_{t_0}$, and compare the result to a thermal state with the energy of the initial state, $\hat \rho_\text{th}$.
We then evaluate $|\overline{\langle\hat{O}_{j}\rangle}_{t_{0}}-\langle\hat{O}_{j}\rangle_{\rm{th}}|$, and the results for different parameters, and correlation functions with different site separations are are summarized in Fig.~\ref{fig:SP_ETH} for spin-1. As implied by Fig.~\ref{fig:SP_ETH_Time}, we again see better agreement with thermal values for short distances and small anisotropy $u$. This relaxation would be expected to be better in a counterflow superfluid regime, and relaxation is less likely to produce the thermal state in the gapped spin-Mott regime, once $u\approx1J$. We do not see a strong transition at this point, which is not surprising for the short-time dynamics observed here. This would be interesting to investigate further in an experimental setting, where we expect that dynamics could be observed over longer times. 

For spin-1/2, we do not plot the results, as the discrepancies from the thermal values $|\overline{\langle\hat{O}_{j}\rangle}_{t_{0}}-\langle\hat{O}_{j}\rangle_{\rm{th}}|\lesssim 0.01$, and the variation with $\Delta$ reflects only differences in the oscillations of the function relative to the averaging time window. With the spin-1/2 model being integrable we might have expected to see deviations from thermal behaviour. However, for these relatively local observables, it seems that the relationship to the energy spectrum is sufficiently simple to allow the short-time values after initial dephasing to reflect the thermal values. Again, it would be interesting to look at this for longer times in experiments, in combination with more complex correlation functions.

Note that for the spin-1/2 case, the possible continuous Luttinger model description~\cite{Giamarchi_Quant_2003} of Hamiltonian \eqref{eq:XYFM_Hamiltonian} makes it possible to observe correlation and thermalization dynamics similar to those in setups with 1D Bose gases~\cite{Gring_Relax_2012, Langen_Exper_2015}. Calculations in those cases thus observe very similar correlation spreading, e.g.~in terms of the evolution of the relative phase of tunnel coupled Bose gases~\cite{Foini_Relax_2017}, displaying very similar results to those in Fig.~\ref{fig:CorrelationEvol_Distance}.

\section{Probing spin currents}
\label{sec:probing_state}

Lastly, we consider how the rotated initial spin states and ground states respond to imposed spin-currents.
In an experiment we can realize this by applying a magnetic field gradient for a short time. 

In each of our models, this corresponds to applying the following kick operator to our states of interest:
\begin{equation}
\hat{\kappa}(\Omega)=\prod_{l}^{M}\me^{-\mi\hat{S}^{z}_{l}{l}\Omega},
\label{eq:kick_op}
\end{equation}
where $\Omega$ denotes a quasi-momentum or ``kick-strength''.
By applying this operator we simulate a short (in time) magnetic field gradient in an experiment, which induces a spin current.
Note, that for the spin-1/2 model, applying operator~\eqref{eq:kick_op} to our initial spin-rotated state is equivalent to preparing ``spin spiral states'' as in recent experiments for studying spin diffusion~\cite{Hild_Far-f_2014}.

We define the spin current $\hat{C}$ as
\begin{equation}
\hat{C} = \frac{1}{M} \sum_l \hat{c}_l,
\end{equation}
with operators
\begin{equation}
\hat{c}_l = - \frac{1}{2\mi} \left( \hat{S}^+_l \hat{S}^-_{l+1} - \hat{S}^-_l \hat{S}^+_{l+1} \right),
\end{equation}
arising from the continuity equation \cite{Transport_Law, SpinCurrents_SF}:
\begin{equation}
\frac{\rm{d}}{\rm{d}t} \hat{S}^z_l = \left[ \mi H, \hat{S}^z_l \right] = -\hat{c}_{l}+\hat{c}_{l-1}.
\end{equation}
Note that in contrast to single particle current measurements \cite{Mun_Phase_2007}, the spin-currents correspond to relative momentum distributions of the two atomic species, and correlations between them could be probed via noise correlation measurements \cite{Altman2004,Greiner2005}.

\begin{figure}[tb] 
\includegraphics[width=\columnwidth]{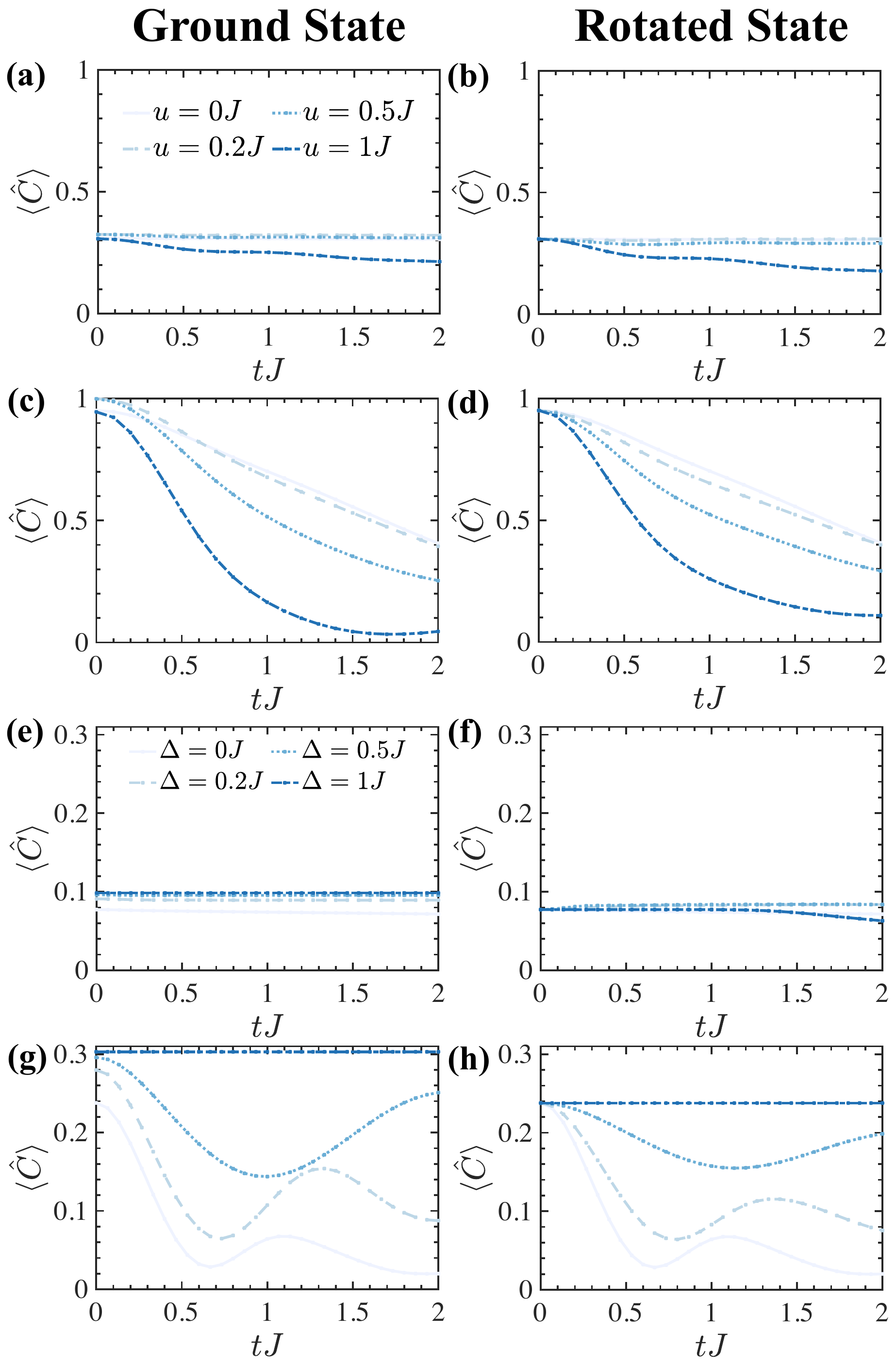}
\caption{
Spin current time evolution after a momentum $\Omega$ is imposed onto the ground state of the Hamiltonian (left column) and the rotated state (right column). The spin current dependence is studied on the anisotropy $u/J$ for spin-1 (top four) and $\Delta/J$ for spin-1/2 (bottom four) cases, after a small $\Omega=0.1\pi$ [panels (a-b/e-f)] and a large $\Omega=0.4\pi$ [panels (c-d/g-h)] momentum kicks. [The numerical calculations were performed for $M=40$ spins with periodic boundary conditions; the numerical convergence was achieved with the MPS bond dimension $D=256$.]
}
\label{fig:SP_Currents_M40_1D_Comparison}
\end{figure}

We compute the currents for spin-1/2 and spin-1, in each case considering the behaviour of the current in the ground state and the rotated state.
We perform calculations by applying the ``kick'' operator as a Matrix Product Operator to the MPS representation of our state, and computing the corresponding time evolution. For an ideal superfluid, we would expect no decay of the current, but interactions will always lead to decay of the current once a critical strength of the kick is exceeded.

The resulting currents are compared in Fig.~\ref{fig:SP_Currents_M40_1D_Comparison}, for both spin-1 and spin-1/2, and for situations where the kick is applied to the ground states (on the left hand side of the figure) or the rotated states (on the right hand side of the figure), for different values of the anisotropies and momenta $\Omega/\pi$.
We can clearly distinguish regimes where the currents are stable and regimes where they are unstable.  Note that for the spin-1/2 case we find that the current imposed to the ground and spin-rotated states both clearly become increasingly stable with increasing $0 \leq \Delta \leq J$, which may be expected as for $\Delta = J$, the Hamiltonian starts to conserve the current. The same result has furthermore been found for the current stability after imposing a ``flux quench'' on interacting spin-less fermions, a model equivalent to our XXZ Heisenberg model~\cite{Nakagawa_Flux_2016}.

Comparing the behaviour of the rotated and ground states, we see qualitatively that up to $tJ=2$ the decay of the current is very similar for both states in the case of each model.
We further observe that the rotated state exhibits a smaller initial current, this is related to a broader initial relative momentum distribution for the two spins (as when the momentum distribution is broader, typically the same translation in quasimomentum will cause less of a change in the average group velocity \cite{Schachenmayer2010}).
Furthermore, we find that in some cases where (especially for spin-1/2 as in Fig.~\ref{fig:SP_Currents_M40_1D_Comparison}(f)) the current is non-decaying for the ground state, we notice a decay for the rotated state with time, as the decay of the long-range correlations (shown in Fig.~\ref{fig:CorrelationEvol_Distance}(d)) becomes important.
The reason why this is particularly visible for spin-1/2 is because the most robust currents occur for larger anisotropy, where there is a bigger difference between the rotated state and the ground state, and hence a faster decay of the correlations.

\begin{figure}[tb] 
\includegraphics[width=\columnwidth]{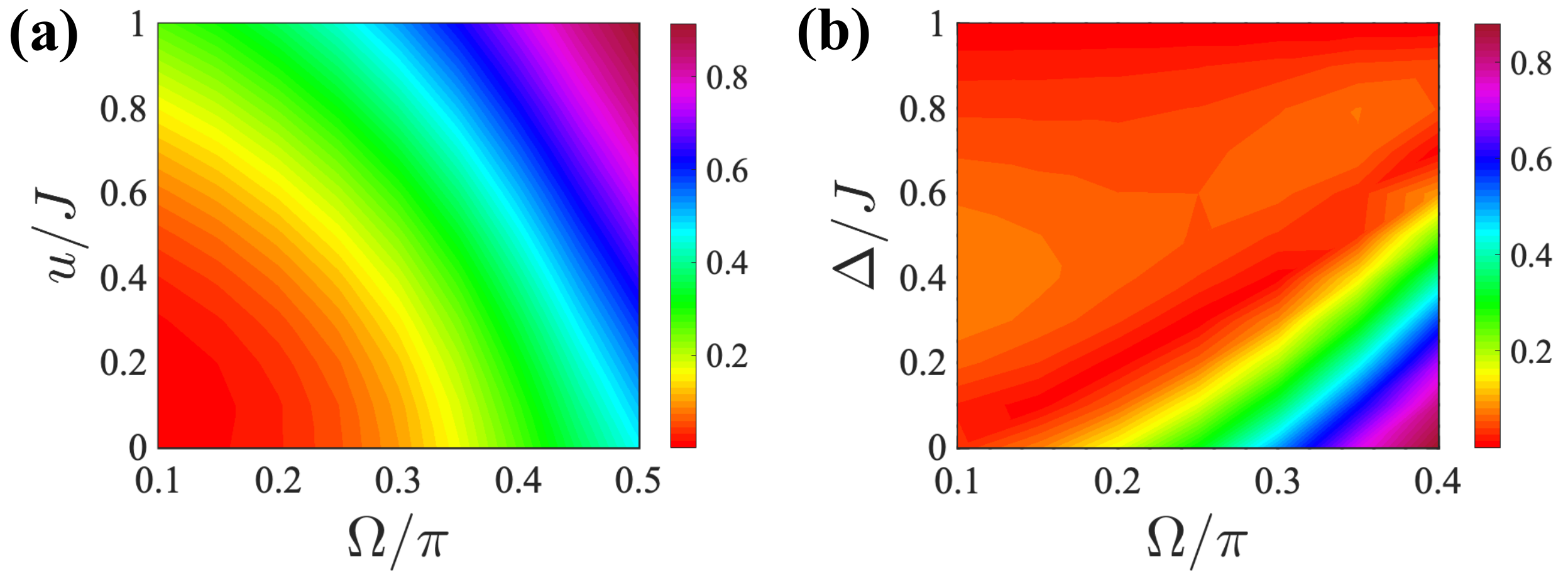}
\caption{
Relative difference $\Delta \braket{C}$ between the spin current at the time $tJ=1$ and at the beginning of the evolution \eqref{eq:Currents_Rel_Dif}.
The figures show the time evolution after a quasi-momentum $\Omega$ of various strength is imposed onto the rotated state for different anisotropies $u/J, \Delta/J$. (a) spin-1, (b) spin-1/2 model.
All calculations are performed for $M=40$, and with periodic boundaries.
The bond dimension used for these MPS calculations was $D=256$.}
\label{fig:SP_Currents_M40_2D_Comparison}
\end{figure}

To emphasize the dependence of the current stability on anisotropy and $\Omega$, in Fig.~\ref{fig:SP_Currents_M40_2D_Comparison} we plot the relative difference between the current after short time evolutions and at the beginning of the evolution, 
\begin{equation}
\Delta \braket{\hat{C}} = \frac{ \bigg| \braket{\hat{C}}_{tJ=1} - \braket{\hat{C}}_{tJ=0^+} \bigg| }{ \braket{\hat{C}}_{tJ=0^+} }.
\label{eq:Currents_Rel_Dif}
\end{equation}
We also find that when the kick is applied to the rotated states, we can clearly quantify a cross-over between two regimes of persistent and decaying currents, in both models.

For the spin-1 model a phase transition occurs for the ground state in the thermodynamic limit between spin superfluid (\textit{XY}) and spin-Mott at $u\approx1J$. We observe the effects of this as a cross-over in the decay rate of currents in our finite-size systems. Specifically, currents rapidly decay as the system becomes more strongly anisotropic, in analogy to spin currents for bosons in a 1D Bose-Hubbard model \cite{Schachenmayer2010,Ketterle_Gradient_Exp,Altman_Lukin_Current3D,Altman_Lukin_Current}. For infinitesimal kicks $\Omega\rightarrow 0$, the currents remain constant in the \textit{XY}-ferromagnetic phase regime, and start to decay once entering into the spin-Mott region. This can be seen along the vertical axis in Fig.~\ref{fig:SP_Currents_M40_2D_Comparison}(a).
For a larger $\Omega$, as we go towards the isotropic point, the current will still decay after a certain critical $\Omega$ value is reached (cross-over value).
This value decreases as we go towards the critical value of $u$ to enter the spin-Mott phase. Again, as in the 1D case for currents in a Bose-Hubbard model \cite{Schachenmayer2010}, this is not a sharp transition, but rather a gradual cross-over, as shown in Fig.~\ref{fig:SP_Currents_M40_2D_Comparison}(a).

In contrast, for spin-1/2, we are always in the \textit{XY}-ferromagnetic phase, where the currents will remain constant for any infinitesimal kick strength $\Omega$, except exactly at the isotropic point.
From \cite{Altman_Lukin_Current3D} we know at the same time that the cross-over value of $\Omega$ increases from zero with increasing anisotropy $\Delta/J$, and we see that the value of $\Omega$ above which we observe substantial decay of the current increases with increasing $\Delta/J$.
For $\Delta=J$, we see essentially non-decaying currents at any time and kick strength from the ground state, which we expect as the $XX$ model can be mapped to non-interacting fermions.

\section{Summary and Outlook}
\label{sec:summary}

For \textit{XY}-ferromagnet states of 1D spin models for two bosonic species in an optical lattice, we have compared and contrasted the dynamics of the ground state and a product state of spins rotated into the \textit{XY} plane, as a function of anisotropy in both spin-1 and spin-1/2 models.
By computing the out-of-equilibrium dynamics, we have shown that in both cases, if we begin in a rotated product spin states, the correlations decrease rapidly in time, faster for a higher anisotropy.
We also compared the rotated state to thermal correlation lengths and entropies.
For the time evolution of spin-currents we observed different behavior between the spin-1/2 and the spin-1 models. For the spin-1/2 model currents are more stable for higher anisotropies, in contrast to the spin-1 case.
This is due to the cross-over velocity in the system increasing with system size.
At longer times, we begin to see decay of currents for the rotated states that occurs earlier than for the ground states, which is where the influence of the correlation decay becomes significant in the dynamics. 
For the spin-1 model, we observed a cross-over between regions where the currents were essentially stable (counterflow superfluid regime, or \textit{XY}-ferromagnet) and unstable (moving towards a spin-Mott state), in analogy to to similar results in superfluid states of 1D Bose-Hubbard models.

By using rf techniques to rotate an initial single-species Mott Insulator state, these states can be directly realized in ongoing experiments with optical lattices.
It is an interesting prospect to probe the difference between mean-field spin states and the true ground states experimentally, for the effective spin models not only in 1D.
For larger dimensions we expect the rotated state to be closer to the true ground-state as the mean-field assumption is generally becoming better with the dimensionality. In particular in this regime, which is hard for fully exact numerical approaches, an experimental investigation would be interesting. 

Using a beyond mean-field formalism for the spin-1/2 model in 3D, there have been predictions for persistent currents after creating spin-spiral states~\cite{Babadi_Far-f_2015}, depending on the selected wave-vector. Also, in 1D, such persistence has been understood due to the integrability of the model~\cite{DeLuca_Noneq_2017,Cominotti_Optim_2014}. It would be interesting to also test the fate of such predictions for the non-integrable spin-1 case both theoretically and experimentally. Generally,
such efforts would provide an interesting basis for further investigation of spin superfluidity in multi-component bosonic lattice models. 

The data for this manuscript is available in open access
at \cite{Strath_Pure}.

\begin{acknowledgments}
    This work was supported by AFOSR MURI Grant No. FA9550-14-1-0035. Work at the University of Strathclyde was supported by the EPSRC Programme Grant DesOEQ (Grant No. EP/P009565/1), and by the EOARD via AFOSR Grant No. FA9550-18-1-0064.
    Numerical calculations here utilized the ARCHIEWeSt High Performance Computer.
    J.S. is supported by the French National Research Agency (ANR) through the Programme d\textsc{\char13}Investissement d\textsc{\char13}Avenir under contract ANR-11-LABX-0058 NIE within the Investissement d\textsc{\char13}Avenir program ANR-10-IDEX-0002-02 and by computational resources of the Centre de calcul de l'Universit\'e de Strasbourg.
    W.K. receives support from the NSF through the Center for Ultracold Atoms and Award No. 1506369, ARO-MURI NonEquilibrium Many-Body Dynamics (Grant No. W911NF14-1-0003), AFOSR-MURI Quantum Phases of Matter (Grant No. FA955014-10035), ONR (Grant No. N00014-17-1-2253), and a Vannevar-Bush Faculty Fellowship. 
\end{acknowledgments}
 
\bibliographystyle{apsrev4-1}
\bibliography{Paper_MIT1}

\end{document}